\documentclass[aps,pra,twocolumn,superscriptaddress,floatfix,letter]{revtex4-1}
\usepackage{epsfig,amsmath,amssymb,color}
\DeclareMathOperator{\sign}{sign}
\usepackage{graphicx}
\usepackage{mathtools}
\usepackage{bm}
\begin{document}

\title{Invisibility cloaks in relativistic motion}
\author{Jad C. Halimeh}
\affiliation{Physics Department and Arnold Sommerfeld Center for Theoretical Physics, Ludwig-Maximilians-Universit\"at M\"unchen, D-80333 M\"unchen, Germany}

\author{Robert T. Thompson}
\affiliation{Department of Mathematics and Statistics, University of Otago, Dunedin 9054, New Zealand}

\author{Martin Wegener}
\affiliation{Institute of Applied Physics and Institute of Nanotechnology, Karlsruhe Institute of Technology (KIT), D-76128 Karlsruhe, Germany}
\date{\today}

\begin{abstract}
We consider an ideal invisibility cloak which is illuminated by monochromatic light and which moves in vacuum at constant relativistic velocity with respect to the common inertial frame of light source and observer. We show that, in general, the moving cloak becomes detectable by image distortions and by generating a broad frequency spectrum of the scattered light. However, for many special combinations of incident light frequency, wave vector of light, and cloak velocity, ideal cloaking remains possible. It becomes non-reciprocal though. This means that light rays emitted by the light source arrive at the observer as though they have travelled through vacuum, but they take completely different paths after being retro-reflected at the observer position.
\end{abstract}

\maketitle

\section{Introduction}
The concept of ideal lossless macroscopic omnidirectional invisibility cloaks operating in free space at one frequency or within a narrow frequency range is consistent with the laws of physics \cite{Sommerfeld1907,Wang2000}. Such spatial cloaks can be designed explicitly by transformation optics \cite{Nicolet1994,Ward1996,Nicolet2004,Leonhardt2006,Pendry2006}. Intuitively, light takes a geometrical detour around an object to be hidden, yet the optical path length (or equivalently the time-of-flight) is conserved despite of the detour. This compensation is accomplished by a suitable material distribution in the cloak containing refractive indices $n=c_0/c<1$, equivalent to a phase velocity $c$ locally exceeding the vacuum speed of light $c_0$. However, for negligible frequency dependence, the energy velocity would be equal to the phase velocity. Thus, as the theory of relativity rules out the possibility of superluminal energy or mass transport \cite{Wald1984}, cloaking devices must unavoidably be frequency dependent (i.e., dispersive) \cite{Monticone2013}. In reality, finite dissipation (i.e., damping) leading to absorption of light imposes further important practical limitations \cite{Hashemi2010}. Nevertheless, it is interesting to consider the theoretical ideal of a loss-less cloak. 

Although electrodynamics is a fully relativistic theory, to date only a few papers have explicitly explored the relativistic nature of transformation optics, including the event cloak \cite{McCall2011} and two studies of cloaks operating in curved spacetimes \cite{Thompson2012jo1,Thompson2015pra}. In this paper, we ask how an ideal cloak would be perceived when moving at constant relativistic velocity with respect to an observer and a monochromatic light source, both at rest with respect to the laboratory inertial frame.

It is well known that setting dielectric media in motion induces magnetoelectric coupling terms \cite{Landau}, so upon first consideration of cloaks in motion one might expect that the cloak must be built to compensate for these induced magnetoelectric terms.  However, this would be incorrect because the stationary observer also perceives a different description of the cloak transformation; one that is stretched out through the motion of the cloak and which results in cloak parameters that correspond exactly to the stationary cloak set in motion. One might subsequently conclude that all concerns about device motion may be ignored during design and construction, but this would also be incorrect. The operational equivalence of a moving cloak to its stationary counterpart is only true for dispersionless cloaks that are not compatible with the laws of physics. 

In this paper, we show that an ideal yet dispersive cloak generally becomes uncloaked by its relative motion (section V), except for infinitely many special cases for which it continues to cloak, but in a non-reciprocal manner (section VI). Before presenting these results, we recall some relevant basics in section II, and define the mathematical basis in sections III and IV.

\section{Moving media and relativistic Doppler shift}
In this section, we briefly recall some aspects of moving media and the relativistic Doppler effect that shall become relevant below. Depending on the angle between the relative velocity vector $\vec{v}$ between two frames of inertia related via a Lorentz transformation from the unprimed (laboratory) coordinates to the primed (co-moving) coordinates, and the wave vector of light $\vec{k}$ in the first and $\vec{k}'$ in the second, one gets the usual longitudinal or transverse relativistic Doppler frequency shift from angular frequency $\omega$ to $\omega'$. For simplicity, we choose the coordinate system such that the relative motion of the two frames is along the $x$-axis. Under these conditions, the Doppler-shifted frequency depends on the angle included by $\vec{v}=(v_x,0,0)^T=c_0(\beta,0,0)^T$ and $\vec{k}=(k_x,k_y,k_z)^T$, and is given by the Lorentz transformation \cite{Misner1973}

\begin{equation}\label{eq:boost}
\begin{pmatrix}
	\omega'/c_0 \\
	\vec{k}'
	\end{pmatrix} = \Lambda
	 \cdot
	\begin{pmatrix}
	\omega/c_0 \\
	\vec{k}
	\end{pmatrix},
\end{equation}

\noindent with

\begin{equation}
	\Lambda=
	\begin{pmatrix}
	\gamma & -\beta\gamma & 0 & 0 \\
	-\beta\gamma & \gamma & 0 & 0 \\
	0 & 0 & 1 & 0 \\
	0 & 0 & 0 & 1		
	\end{pmatrix}
\end{equation}

\noindent and with the Lorentz factor

\begin{equation}
\gamma=\frac{1}{\sqrt{1-\beta^2}}.
\end{equation}

This frequency shift also applies to the dipoles in the material forming the cloak: The dipoles receive the Doppler-shifted incoming light wave and re-emit a wave with the same frequency in the co-moving frame. Due to the relative motion with respect to the observer, a second Doppler shift occurs. This second shift compensates the first one if the direction of light impinging onto the medium is parallel to the direction of light emerging from it. A parallel dielectric plate moving in vacuum is a simple example. The situation is different if the emerging wave vector of light includes with the incident wave vector of light in the co-moving frame a non-zero deflection characterized by the general intrinsic rotation $R=R_{z'}(\phi)R_{y'}(\theta)R_{x'}(\psi)$, with Tait-Bryan angles $\phi$, $\theta$, and $\psi$ about the $z'$-, $y'$-, and $x'$-axis, respectively, in which case one does get a net Doppler frequency shift. If we denote by $\tilde{\omega}$ the laboratory-frame frequency of the emerging wave vector of light, the relative frequency shift $\Delta\omega/\omega\equiv(\tilde{\omega}-\omega)/\omega$ for the longitudinal case is given by

\begin{equation}\label{eq:freqshiftlong}
\left(\frac{\Delta\omega}{\omega}\right)_{\text{long}}=\frac{\beta}{\beta+\zeta}\left(\cos\phi\cos\theta-1\right),
\end{equation}

\noindent where $\zeta=\sign(\vec{k}\cdot\hat{x})$. For the transverse case the relationship is more complicated, and, unlike the longitudinal case, depends additionally on $\psi$: 

\begin{multline}\label{eq:freqshifttrans}
\left(\frac{\Delta\omega}{\omega}\right)_{\text{trans}}=\beta\gamma[\beta\gamma-\xi\cos\psi\sin\phi+\\
\left(\xi\sin\psi\sin\theta-\beta\gamma\cos\theta\right)\cos\phi],
\end{multline}

\noindent where $\xi=\sign(\vec{k}\cdot\hat{y})$.

A simple example is a slightly wedged dielectric plate moving with constant velocity perpendicular to the line of sight between light source and observer. Non-relativistically, one can argue that the moving wedge linearly increases or decreases the amount of material in the optical path versus time, leading to a time-dependent phase shift of the light wave, hence to a frequency and momentum shift of the ÒtransmittedÓ light. Correspondingly, the quanta of light have a smaller or larger photon energy and momentum compared to the wedge at rest with respect to the laboratory frame. The energy difference is taken from or added to the kinetic energy of the wedge. Likewise, the change in momentum of light is compensated by the wedge. However, for macroscopic objects and small light intensities, we can safely neglect this back-action onto the moving object.

\section{Dispersive cylindrical cloak}
Our mathematical treatment starts with the spatial distribution of the electric permittivity $\overset\leftrightarrow{\epsilon}$ and magnetic permeability $\overset\leftrightarrow{\mu}=\overset\leftrightarrow{\epsilon}$ tensors for the cloak operation frequency $\omega_0$ obtained from Ref.~\onlinecite{Pendry2006} for the transformation of a line to a cylinder with radius $R_1$ using the simple linear mapping (in cylindrical coordinates) 

\begin{equation}
r\rightarrow r'=R_1+\frac{R_2-R_1}{R_2}r
\end{equation}

\noindent for $0\leq r\leq R_2$, with the outer radius of the cloaking shell $R_2$. The other coordinates remain untransformed. This yields the optical tensors $\overset\leftrightarrow{\epsilon} = \overset\leftrightarrow{\mu}=\overset\leftrightarrow{n}$ given by \cite{Pendry2006,Schurig2006}

\begin{equation}
\overset\leftrightarrow{\epsilon} =
         \begin{pmatrix}
	\epsilon_{r'}(r',\omega_0) & 0 & 0 \\
	0 & \epsilon_{\theta'}(r',\omega_0) & 0 \\
	0 & 0 & \epsilon_{z'}(r',\omega_0)
	\end{pmatrix}
\end{equation}

\noindent with

\begin{align}
\epsilon_{r'}(r',\omega_0) &=\frac{r'-R_1}{r'}, \\
\epsilon_{\theta'}(r',\omega_0) &=  \frac{r'}{r'-R_1}, \\
\epsilon_{z'}(r',\omega_0) &=  \left(\frac{R_2}{R_2-R_1}\right)^2 \frac{r'-R_1}{r'},
\end{align}

\noindent where $r'\in[R_1,R_2]$ represents here the radial coordinate of light, $R_1$ ($R_2$) is the inner (outer) radius of the cylindrical shell comprising the cloak. It is straightforward to check that for our choice of $R_2=2R_1$, as $r':R_2\rightarrow R_1$, then $\epsilon_{r'}(r',\omega_0):0.5\rightarrow 0$, $\epsilon_{\theta'}(r',\omega_0): 2\rightarrow +\infty$, and $\epsilon_{z'}(r',\omega_0):2\rightarrow 0$.

When the light has a frequency $\omega'$ that is not necessarily equal to $\omega_0$, due to the inherently dispersive nature \cite{Monticone2013} of the cloak, the optical tensor component $\epsilon_{i'}(r',\omega')$ will no longer equal $\epsilon_{i'}(r',\omega_0)$ ($i'=r',\theta',z')$. To describe this aspect, we use an undamped Lorentz oscillator model for the material dispersion. It is often expressed by some constant background permittivity plus the response of a harmonic oscillator. Microscopically, the background contribution stems from other resonances at (much) higher frequencies. To avoid unphysical frequency-independent contributions, we model the background by a second harmonic oscillator. In the co-moving frame, suppressing the tensor indices for clarity, we get for any of the tensor components the form

\begin{multline}
\epsilon(r',\omega')=\mu(r',\omega')=\\
1+\frac{f_1}{\Omega_1^2(r')-\omega'^2}+\frac{f_2}{\Omega_2^2(r')-\omega'^2},
\end{multline}

\noindent where the varying eigenfrequencies are related by $\Omega_2=\alpha\Omega_1$. The space-dependent resonance frequency $\Omega_1(r')$ is computed by honoring the condition that  the tensor components $\epsilon(r',\omega=\omega_0)=\epsilon_{\rm ideal}$ satisfy the ideal cloak parameters. This allows for a quadratic equation ${\cal A}X^2+{\cal B}X+{\cal C}=0$, where $X=\Omega_1^2$ and 

\begin{align}
{\cal A} &= (\epsilon_{\rm ideal}-1)\alpha^2, \\
{\cal B} &= -\left[(1+\alpha^2)(\epsilon_{\rm ideal}-1)\omega_0^2+f_1\alpha^2+f_2\right], \\
{\cal C} &= \left[(\epsilon_{\rm ideal}-1)\omega_0^2+f_1+f_2\right]\omega_0^2.
\end{align}

\noindent For our choice of parameters $f_1=\omega_0^2$, $f_2=10\,\omega_0^2$, and $\alpha=10$, we find non-negative $X=(-{\cal B}-\sqrt{{\cal B}^2-4{\cal A}{\cal C}})/2{\cal A}$ solutions for all $r'$, from which we then calculate $\Omega_1=\sqrt{X}$. The choice of the oscillator parameters is obviously not unique, but none of the qualitative results discussed below depends on the particular parameter choices. Furthermore, we add an ideal metallic mirror at the inner shell with radius $R_1$ to isolate the interior from the outside.

\section{Ray tracing approach}
Images of various non-moving cloaks have previously been calculated by advanced ray-tracing \cite{Schurig2006,Danner2010a,Danner2010b,Halimeh2011,Halimeh2012a}. The images of ordinary objects (i.e., not cloaks) moving at relativistic velocities or even subject to acceleration/gravitation have been simulated extensively as well \cite{Ruder2008,James2015a,James2015b}. Various image distortions arise due to the fact that an observer sees light rays arriving at one point in time in his/her system of inertia, while these light rays may have been emitted from the moving object at different points in time as seen from the co-moving frame. 

All ray-tracing calculations presented in this paper are performed in a frame co-moving with the cloak and are based on the Hamiltonian formulation for light propagation outlined in Ref.~\onlinecite{Schurig2006} for a general (i.e., possibly dispersive) impedance-matched medium.  We use their \cite{Schurig2006} nomenclature in this section. By replacing $\vec{x} \rightarrow \vec{r}'$ and $\vec{k}\rightarrow \vec{k}'$, one arrives at the nomenclature used in all other sections of our paper. From Maxwell's curl equations and the parameter relations $\vec{D}=\epsilon_0\overset\leftrightarrow{\epsilon}\vec{E}$ and $\vec{B}=\mu_0\overset\leftrightarrow{\mu}\vec{H}$, we get the relation \cite{Schurig2006}

\begin{equation}
\vec{k}\times\left[\overset\leftrightarrow{\epsilon}^{-1}\left(\vec{k}\times\vec{E}\right)\right]+\overset\leftrightarrow{\epsilon}\vec{E}=0
\end{equation}

\noindent that requires, for non-zero field solutions, that

\begin{equation}
\frac{1}{|\overset\leftrightarrow{\epsilon}|}\left(\vec{k}^T\overset\leftrightarrow{\epsilon}\vec{k}-|\overset\leftrightarrow{\epsilon}|\right)^2=0,
\end{equation}

\noindent from which we obtain the Hamiltonian $\mathcal{H}=\vec{k^T}\overset\leftrightarrow{\epsilon}\vec{k}-|\overset\leftrightarrow{\epsilon}|$. The ray tracing is then performed by iterative integration of Hamilton's equations

\begin{align}
\frac{d\vec{x}}{d\tau}&=\frac{\partial \mathcal{H}}{\partial \vec{k}}, \\
\frac{d\vec{k}}{d\tau}&=-\frac{\partial \mathcal{H}}{\partial \vec{x}}, 
\end{align}

\noindent with $\vec{x}$ as spatial position of light, and $\tau$ is a parametrization variable that can represent time, but not necessarily real time. 

Note that the Hamiltonian used here allows us to consistently treat positive and negative refractive indices. The latter do occur under the conditions of moving cloaks below.

At the boundary between vacuum and cloak, the refractive index generally makes a discontinuous jump. We treat this problem by ordinary refraction, including the possibility of total internal reflection. Ordinary Fresnel reflections do not occur at the vacuum-cloak boundary because even the moving cloak is always perfectly impedance matched to vacuum by the condition $\overset\leftrightarrow{\mu}=\overset\leftrightarrow{\epsilon}$.

\section{Uncloaking by motion}
It is instructive to ``illegally'' ignore material dispersion for a moment. As already pointed out in the introduction, in the laboratory frame, the magnetoelectric material distribution of the cloak at rest leads to a complicated bi-anisotropic material distribution of the moving cloak \cite{Thompson2011}. In this frame, the result is not obvious. The situation is simpler in the co-moving frame: A light ray emerging from the source in the laboratory frame can be Lorentz-transformed into the frame co-moving with the cloak. In this frame, the cloak is electromagnetically equivalent to vacuum. Hence, light rays outside the cloak propagate as in vacuum. After passing the cloaking structure, the rays can be Lorentz-transformed back to the observer frame. This means that the constant motion of the cloak has no effect at all on the cloaking performance.

\begin{figure}[]
\includegraphics{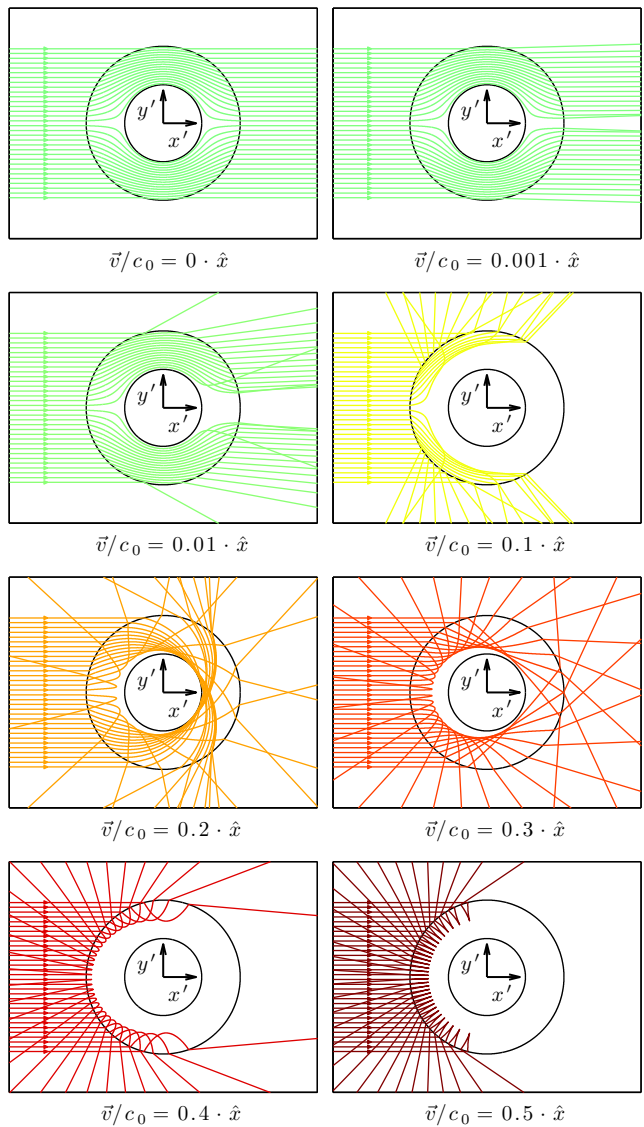}
\caption{(Color online)
Characteristic light rays in a frame ($x' y'$-plane) co-moving with a dispersive cylindrical invisibility cloak moving with constant velocity $v$ along the positive $x$-direction in the laboratory system of light source and observer. The cloak radii $R_1$ and $R_2$ are indicated by the black circles, and ray colors indicate light frequency with $\omega_0$ as green. The invariance of the cylindrical cloak along the $z'$-axis allows all optics to be captured in the $x' y'$-plane. The monochromatic light rays with frequency $\omega=\omega_0$ are emitted along the positive $x$-direction in the laboratory frame. $\omega_0$ is the cloak operation frequency, i.e., if the cloak is at rest with respect to the light source, the condition $\omega=\omega_0$ leads to perfect cloaking. With increasing velocity $\beta=v/c_0$ with respect to the vacuum speed of light $c_0$, increasing aberrations are found. Light rays for which the emerging direction (or wave vector of light $\vec{k}'$) includes an angle $\phi$ with the incident rays in the $x'y'$-plane are received with a frequency $\tilde{\omega}\neq\omega$ by an observer at rest in the laboratory frame. This leads to the relative frequency shift $\Delta\omega/\omega\equiv(\tilde{\omega}-\omega)/\omega=(\cos\phi-1)\beta/(\beta+1)$, a simplified version of Eq.~\eqref{eq:freqshiftlong} for the geometry shown here.
}
\label{fig:Fig1}
\end{figure}

The behavior is different for a dispersive cloak. Suppose we illuminate with a frequency of light $\omega$ equal to the cloak operation frequency $\omega_0$ in the laboratory frame. Upon Doppler frequency shifting from the light source to a frame co-moving with the cloak, $\omega\rightarrow\omega'$, the frequency of light  is in general no longer equal to the operation frequency $\omega_0$. As a result, the spatial distribution of electric permittivity and magnetic permeability as seen from the cloak frame is different from that of the ideal cloak. Hence, the cloak generally exhibits aberrations. 

A notable exception is the case where the light frequency in the laboratory frame is pre-compensated such that it equals the cloak operation frequency after being Doppler shifted to the cloak frame. For example, if the cloak moves towards the observer along the axis connecting source and observer, i.e., away from the light source, the frequency emitted from the light source needs to be blue-shifted to compensate the longitudinal Doppler red shift. In this case, light rays propagate from the light source to the observer as though there was vacuum in between. In particular, light reaches the observer with the same frequency as emitted by the source. Cloaking is conceptually ideal in this case, but it is fundamentally non-reciprocal: If a light ray is retro-reflected by a mirror at the observer location, the cloak is moving towards this new source. Hence, the original pre-compensation towards the blue and the longitudinal Doppler blue shift add up (rather than subtract), leading to even larger aberrations of the cloak. We will come back to the possibility of pre-compensation and non-reciprocal cloaking in section VI.

\begin{figure}[]
\includegraphics[width=1.\linewidth]{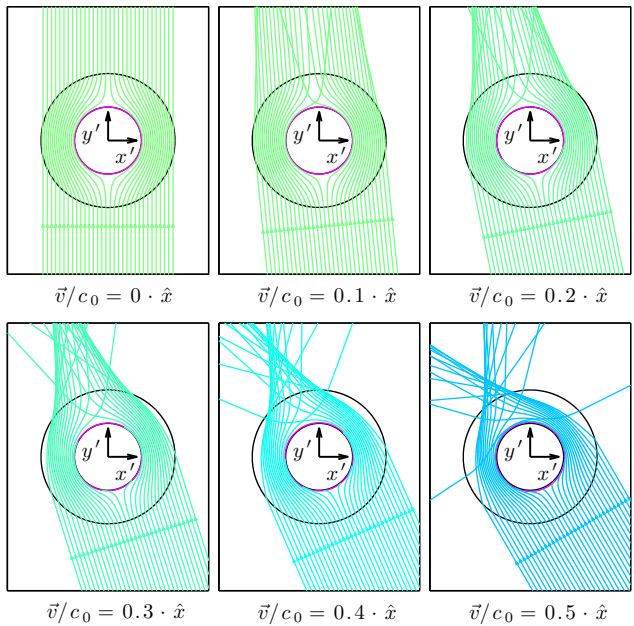}
\caption{(Color online)
As Fig.~\ref{fig:Fig1}, but for light rays with $\omega=\omega_0$ impinging along the positive $y$-direction (rather than the positive $x$-direction) in the laboratory frame. In the shown co-moving frame, the incident rays at the bottom are inclined with respect to the $y'$-axis due to the Lorentz transformation, although they are along the positive $y$-direction in the laboratory frame. Again, with increasing cloak velocity $v/c_0$, more pronounced aberrations are observed. Some rays end within the cloak because they hit the vicinity of local resonances at a radius indicated by the magenta-color circle. For arbitrarily small yet non-zero damping, these rays will be absorbed. As in the longitudinal case, there will also be a frequency spread due to deflections in the direction of the wave vectors with respect to their incidence directions upon interacting with the cloak.
}
\label{fig:Fig2}
\end{figure}

To get a better feeling for what kind of aberrations can be expected for cloaks moving at finite velocities, we depict various examples for calculated light-ray paths in Fig.~\ref{fig:Fig1} for a source shining light of frequency $\omega=\omega_0$ along the positive $x$-direction.  The ratio $v/c_0$ of cloak laboratory-frame velocity $v$ to the vacuum speed of light $c_0$ increases as indicated. Positive velocities $v>0$ mean that the cloak is moving in the positive $x$-direction away from the light source, i.e., $\omega'<\omega$. For up to $v/c_0=10^{-4}$, no distortions of the light rays are visible within the linewidths of the plots. At velocities approaching about $v/c_0=1\%$, the distortions with respect to parallel ray paths become quite prominent. At $v/c_0=10\%$, some rays already emerge under large angles with respect to their direction of incidence and a few are totally-internally reflected at the vacuum-cloak interface, never entering the cloak. At $v/c_0=20\%$, some rays are reflected inside the cloak at the inner shell, which we have chosen to be perfect mirror (see above). Upon Lorentz transformation of these directions (i.e., wave vectors $\vec{k}'$) back to the laboratory frame, the frequency of light corresponding to these rays gets shifted to $\tilde{\omega}$ (see introductory discussion above). The spread in angles thus leads to a spread in relative frequency shifts $\Delta\omega/\omega=(\tilde{\omega}-\omega)/\omega$ given in Eq.~\eqref{eq:freqshiftlong}. Obviously, both aspects, the change in light direction with respect to vacuum and the frequency shift with respect to the incident monochromatic light make the cloak detectable.

The behavior for light rays impinging orthogonal to the cloak motion as seen from the laboratory frame is depicted in Fig.~\ref{fig:Fig2}. Due to the Lorentz transformation, orthogonal in the laboratory frame is no longer orthogonal in the co-moving frame. With increasing cloak velocity, we again find increasing distortions like in the parallel case (Fig.~\ref{fig:Fig1}). Some rays in Fig.~\ref{fig:Fig2} arrive at locations for which $\omega'>\omega=\omega_0$ approaches a local resonance. The azimuthal component of the refractive index diverges and, hence, these rays asymptotically approach an orbit at some radius, in the vicinity of which we end these rays. Physically, they are absorbed at this radius as soon as one introduces an arbitrarily small yet non-zero damping. In general, light rays in the moving cloak also experience regions with negative permittivity and permeability, hence negative phase velocity, leading to negative refraction.

\begin{figure}[]
\includegraphics[width=1.\linewidth]{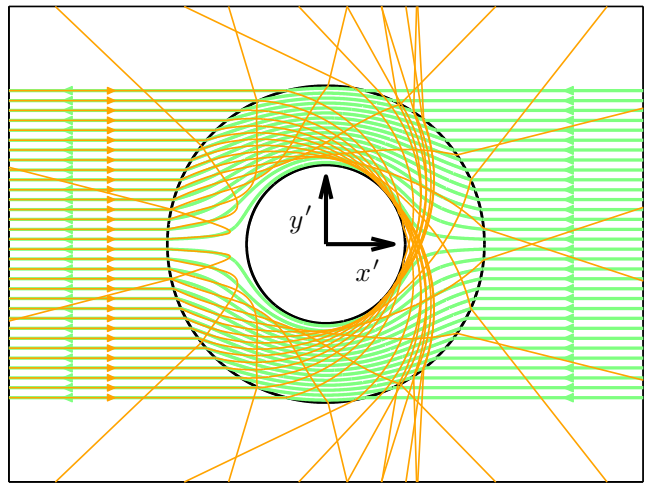}
\caption{(Color online)
As in Fig.~\ref{fig:Fig1}, but for the light frequency $\omega=\omega_0\sqrt{(1-\beta)/(1+\beta)}$ in the laboratory frame with $\beta=v/c_0=0.1$ such that the Doppler blue-shifted frequency in the co-moving frame obeys $\omega'=\omega_0$, leading to perfect cloaking (green color) for light propagating with frequency $\omega$ along the negative $x$-direction in the laboratory frame. In sharp contrast, upon retro-reflecting these rays at the observer (where again $\omega=\omega_0\sqrt{(1-\beta)/(1+\beta)}$) strong aberrations are observed because $\omega'=\omega_0(1-\beta)/(1+\beta)$ due to a Doppler red shift from the laboratory to the co-moving frame. This means that the cloaking behavior is nonreciprocal. Infinitely many such special configurations exist for the different possible laboratory-frame frequencies $\omega$.
}
\label{fig:Fig3}
\end{figure}

\section{Non-reciprocal cloaking}
In the previous section, we have emphasized that motion at relativistic velocities generally reveals the cloak. However, for any cloak velocity, one can find incident wave vectors of light $\vec{k}=(k_x,k_y,k_z)^T$ and a frequency of light $\omega$ connected via Eq.~\eqref{eq:boost} by 

\begin{equation}\label{eq:nrc}
\omega'=\gamma(\omega-vk_x)=\omega_0
\end{equation}

\noindent such that the Doppler shift leads to a frequency $\omega'=\omega_0$ in the co-moving frame equal to the cloak operation frequency $\omega_0$. Together with the vacuum dispersion relation of light $k_x^2+k_y^2+k_z^2=\omega^2/c_0^2$, the wave vectors of light obeying this condition lie on a cone. All light rays impinging along these directions arrive at the observer as though they have propagated in vacuum. However, upon retro-reflecting these rays at the observer, i.e., upon replacing $\vec{k}\rightarrow-\vec{k}$, we get $\omega'\neq\omega_0$. This means that no cloaking is obtained. Altogether, cloaking is non-reciprocal. This behavior is illustrated in Fig.~\ref{fig:Fig3} for a cloak with velocity $\beta=v/c_0=10\%$ in the positive $x$-direction and light propagating along the $x$-axis in the laboratory frame with a frequency $\omega=\omega_0\sqrt{(1-\beta)/(1+\beta)}$. 

\begin{figure}[]
\includegraphics[width=0.8\linewidth]{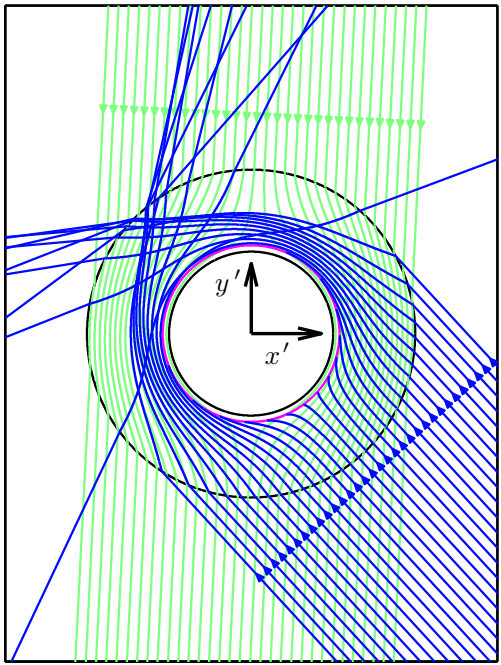}
\caption{(Color online)
Non-reciprocal cloaking for light with frequency $\omega=\omega_0$ (green rays) and wave vector $\vec{k}$ in the laboratory frame obeying Eq.~\eqref{eq:nrc} and impinging on the cylindrical invisibility cloak while the latter is moving at relative velocity $v/c_0=0.5$ in the positive $x$-direction. This light has the same frequency $\omega'=\omega_0$ in the co-moving frame as well. Hence, the cloaks works perfectly. However, upon retro-reflection of the light in the laboratory frame, its wave vector becomes $-\vec{k}$, and this light will no longer have a frequency $\omega'=\omega_0$ in the co-moving frame as it acquires a Doppler blue-shift, leading to imperfect cloaking. A few rays end within the cloak as they hit the vicinity of a local resonance at a radius indicated by the magenta circle. For arbitrarily small yet non-zero damping, these rays are absorbed there. Finally, note that even though $\vec{k}$ and $-\vec{k}$ are retro-reflections of one another in the laboratory frame, their Lorentz boosts in the co-moving frame are not. In other words, if $\vec{k}$ is boosted to $\vec{k}'$, that does not mean that $-\vec{k}$ will get boosted to $-\vec{k}'$, i.e., retro-reflection in the laboratory frame appears a bit unintuitive if represented in the co-moving frame.
}
\label{fig:Fig4}
\end{figure}

Alternatively, if the light in the laboratory frame has the operational frequency $\omega=\omega_0$ of the cloak, one can find the proper direction in the laboratory frame that this light has to take as per Eq.~\eqref{eq:nrc} so that $\omega'=\omega_0$. The cloak works perfectly for these directions but not for the opposite ones. This is illustrated in Fig.~\ref{fig:Fig4} for the example of a cloak with velocity of $\beta=v/c_0=50\%$ in the positive $x$-direction. As the incident direction of light is not parallel to the direction of motion in Fig.~\ref{fig:Fig4} (in contrast to Fig.~\ref{fig:Fig3}), retro-reflection of light in the laboratory frame according to $\vec{k}\rightarrow-\vec{k}$ does not correspond to $\vec{k}'\rightarrow-\vec{k}'$ in the co-moving frame. The wave vectors $\vec{k}'$ rather follow from Eq.~\eqref{eq:boost}.

\section{Conclusion}
In conclusion, we have discussed the behavior of relativistically moving cylindrical cloaks in vacuum. We have shown that an unavoidably dispersive cloak is generally uncloaked due to its motion and the resulting relativistic Doppler frequency shift except for many special cases for which cloaking remains perfect but becomes non-reciprocal. Broadly speaking, these results show that a physical invisibility cloak is not equivalent to ordinary electromagnetic vacuum -- not even at a single frequency. 

We have focused on the example of a cylindrical cloak designed by a linear spatial transformation throughout this article. However, it should be clear by our general discussion that the findings outlined in this article apply to other free-space electromagnetic cloaks moving with constant velocity as well. In accelerated frames or, equivalently, under the influence of gravity, any dispersive free-space cloak designed along the lines of \cite{Leonhardt2006,Pendry2006} for flat space-time will no longer work perfectly either, because the local frequency of light generally changes due to gravity \cite{Wald1984}.

\section{Acknowledgments}
J.C.H. acknowledges stimulating discussions with Thomas Udem (MPQ Garching), Michael Haack, and Andreas Weichselbaum (LMU Munich), and extends his gratitude to Husni Habal (TU Munich) for help with the esthetic arrangement of the Figures. R.T.T. is supported by the Royal Society of New Zealand through Marsden Fund Fast Start Grant No. UOO1219.



\begin{thebibliography}{9}

\bibitem{Sommerfeld1907}
A. Sommerfeld,
\emph{Ein Einwand gegen die Relativtheorie der Elektrodynamik und seine Beseitigung}.
Physikalische Zeitschrift \textbf{8}(23), 841-842 (1907).

\bibitem{Wang2000}
L. J. Wang, A. Kuzmich, and A. Dogariu,
\emph{Gain-assisted superluminal light propagation}.
Nature \textbf{406}, 277-279 (2000).

\bibitem{Nicolet1994}
A. Nicolet, J.-F. Remacle, B. Meys, A. Genon, and W. Legros,
\emph{Transformation methods in computational electromagnetism}.
J. Appl. Phys. \textbf{75}(10), 6036-6038 (1994).

\bibitem{Ward1996}
A.J. Ward and J.B. Pendry, 
\emph{Refraction and geometry in MaxwellÕs equations}.  
J. Mod. Opt. 43 (4), 773-793 (1996).

\bibitem{Nicolet2004}
A. Nicolet, S. Guenneau, and F. Zolla,
\emph{Modelling of twisted optical waveguides with edge elements}.
Eur. Phys. J. Appl. Phys. \textbf{2}(28), 153-157 (2004).

\bibitem{Leonhardt2006}
U. Leonhardt,
\emph{Optical Conformal Mapping}.
Science \textbf{312}, 1777-1780 (2006).

\bibitem{Pendry2006}
J. B. Pendry, D. Schurig, and D. R. Smith, 
\emph{Controlling Electromagnetic Fields}.
Science \textbf{312}, 1780-1782 (2006).

\bibitem{Wald1984}
R. M. Wald,
\emph{General Relativity}.
(The University of Chicago Press, 1984).

\bibitem{Monticone2013}
F. Monticone and A. Al\`u,
\emph{Do Cloaked Objects Really Scatter Less?}.
Phys. Rev. X \emph{3}, 041005 (2013).

\bibitem{Hashemi2010}
H. Hashemi, B. Zhang, J. D. Joannopoulos, S. G. Johnson,
\emph{Delay-Bandwidth and Delay-Loss Limitations for Cloaking of Large Objects}.
Phys. Rev. Lett. \textbf{104}, 253903 (2010).

\bibitem{McCall2011}
M. W. McCall, A. Favaro, P. Kinsler, and A. Boardman,
\emph{A spacetime cloak, or a history editor}.
J. Opt. \textbf{13}(2), 024003 (2011).

\bibitem{Thompson2012jo1}
R. T. Thompson,
\emph{General relativistic contributions in transformation optics}.
J. Opt. \textbf{14}(1), 015102 (2012).

\bibitem{Thompson2015pra}
R. T. Thompson and M. Fathi,
\emph{Shrinking cloaks in expanding space-times: The role of coordinates and the meaning of transformations in transformation optics}.
Phys. Rev. A \textbf{92}(1), 013834 (2015).

\bibitem{Landau}
L. D. Landau, E. M. Lifshitz, and L. P. Pitaevskii,
\emph{Electrodynamics of Continuous Media,
Landau and Lifshitz Course of Theoretical Physics VIII (2nd. ed.)}.
(Elsevier,1984).

\bibitem{Misner1973}
C. W. Misner, K. S. Thorne, and J. A. Wheeler,
\emph{Gravitation}.
(Macmillan, 1973).

\bibitem{Schurig2006}
D. Schurig, J. B. Pendry, and D. R. Smith,
\emph{Calculation of material properties and ray tracing in transformation media}.
Opt. Express \textbf{14}(21), 9794-9804 (2006).

\bibitem{Danner2010a}
A. J. Danner,
\emph{Visualizing invisibility: metamaterials-based optical devices in natural environments}. 
Opt. Express \textbf{18}(4), 3332-3337 (2010).

\bibitem{Danner2010b}
A. Akbarzadeh and A. J. Danner,
\emph{Generalization of ray tracing in a linear inhomogeneous anisotropic medium: a coordinate-free approach}. 
J. Opt. Soc. Am. A \textbf{27}, 2558-2562 (2010).

\bibitem{Halimeh2011}
J. C. Halimeh, R. Schmied, and M. Wegener,
\emph{Newtonian photorealistic ray tracing of grating cloaks and correlation-function-based cloaking-quality assessment}.
Opt. Express \textbf{19}(7), 6078-6092 (2011).

\bibitem{Halimeh2012a}
J. C. Halimeh and M. Wegener,
\emph{Time-of-flight imaging of invisibility cloaks}.
Opt. Express \textbf{20}(1), 63-74 (2012).

\bibitem{Ruder2008}
H. Ruder, D. Weiskopf, H.-P. Nollert, and T. M\"uller,
\emph{How computers can help us in creating an intuitive access to relativity}.
New J. Phys. \textbf{10}, 125014 (2008).

\bibitem{James2015a}
O. James, E. von Tunzelmann, P. Franklin, and K. S. Thorne,
\emph{Gravitational lensing by spinning black holes in astrophysics, and in the movie Interstellar}.
Classical and Quantum Gravity \textbf{32}, 065001 (2015).

\bibitem{James2015b}
O. James, E. von Tunzelmann, P. Franklin, and K. S. Thorne,
\emph{Visualizing Interstellar's Wormhole}.
Am. J. Phys. \textbf{83}, 486-499 (2015)

\bibitem{Thompson2011}
R. T. Thompson, S. A. Cummer, and  J. Frauendiener,
\emph{A completely covariant approach to transformation optics}.
J. Opt. \textbf{13}(2), 024008 (2011).

\end{thebibliography}
\end{document}